\documentstyle[11pt,epsf]{article}
 1
 1
 1

\def\lsim{\mathrel{\raise.3ex\hbox{$<$\kern-.75em\lower1ex\hbox{$\sim$}}}}
\def\gsim{\mathrel{\raise.3ex\hbox{$>$\kern-.75em\lower1ex\hbox{$\sim$}}}}
\begin{document}
\noindent
\thispagestyle{empty}
\renewcommand{\thefootnote}{\fnsymbol{footnote}}
\begin{flushright}
{\bf UCSD/PTH 97-09}\\
{\bf hep-ph/9704325}\\
{\bf April 1997}\\
\end{flushright}
\vspace{.5cm}
\begin{center}
  \begin{Large}\bf
Perturbative ${\cal{O}}(\alpha_s^2)$ Corrections to the Hadronic Cross
Section  \\[2mm]
Near Heavy Quark-Antiquark Thresholds in $e^+e^-$ Annihilation
  \end{Large}
  \vspace{1.5cm}

\begin{large}
 A.H. Hoang
\end{large}
\begin{center}
\begin{it}
   Department of Physics,
   University of California, San Diego,\\
   La Jolla, CA 92093--0319, USA\\ 
\end{it} 
\end{center}

  \vspace{4cm}
  {\bf Abstract}\\
\vspace{0.3cm}

\noindent
\begin{minipage}{15.0cm}
\begin{small}
It is demonstrated how perturbative ${\cal{O}}(\alpha_s^2)$ 
relativistic corrections to the non-relativistic stable 
heavy quark-antiquark production cross section in $e^+e^-$ 
annihilation based on a Coulombic QCD potential can
be systematically calculated using the concept of effective
field theories. The ${\cal{O}}(\alpha_s^2)$ corrections from the 
relativistic energy-momentum relation, the relativistic phase space
corrections and the ${\cal{O}}(\alpha_s^2)$ corrections involving the
group theoretical factors $C_F^2$ and $C_F T$ are determined
explicitly. For the case of $t\bar t$ production the sum of these 
corrections amounts to  $3\%$ -- $7\%$ over the whole threshold regime
and is insensitive to variations in the top width.
Perturbative ${\cal{O}}(\alpha_s^2)$ corrections to the leptonic decay
width of heavy quark-antiquark ${}^3S_1$ vector resonances are
extracted. 
\end{small}
\end{minipage}
\end{center}
\setcounter{footnote}{0}
\renewcommand{\thefootnote}{\arabic{footnote}}
\vspace{1.2cm}
%
%
%
\newpage
\noindent
In the kinematic regime where the c.m. energy is much larger than the
quark masses the total hadronic cross section in $e^+e^-$ collisions
$R_{had} = \sigma(e^+e^-\to\mbox{Hadrons})/\sigma_{pt}$, where 
$\sigma_{pt}$ is the point cross section, belongs to the theoretically
best known and understood quantities in electroweak physics. Inspired
by the asthonishing precision of the LEP experiments $R_{had}$ has
been calculated up to ${\cal{O}}(\alpha_s^3)$ accuracy for LEP
energies based on sophisticated multi-loop techniques~\cite{multiloop}. 
In the kinematic regime near quark-antiquark thresholds, however, which
is characterized by the relation 
\begin{equation}
|\beta|\,\lsim\,\alpha_s
\,,\qquad
\beta\,\equiv\,\sqrt{1-\frac{4M_Q^2}{s+i\epsilon}}
\,,
\end{equation}
where $M_Q$ is the quark mass\footnote{
Throughout this paper the quark mass $M_Q$ is understood as the pole
mass and $\beta$ is called ``velocity''.
}
and $\sqrt{s}$ the c.m. energy,
the theoretical understanding of $R_{had}$ is much poorer. Although 
perturbative methods to describe the hadronic cross section in the 
threshold regime in general seem to be more than questionable due to 
non-perturbative effects and due to the large size of the strong 
coupling, it is well accepted that a perturbation theory based 
description of the hadronic cross section near a quark-antiquark
threshold should be possible if the quark mass is sufficiently larger 
than the hadronic scale $\Lambda_{\mbox{\tiny QCD}}$~\cite{Billoire1}. 
The prototype
application of perturbative methods in the threshold regime is therefore
the case of $t \bar t$ production\footnote{
An application of perturbative methods might also be possible for
the $b \bar b$ production cross section 
for low radial excitations of the $\Upsilon$ family 
because in that case non-perturbative effects seem to be well under 
control~\cite{Yndurain1,Yndurain2} and the strong coupling is still small 
enough that a perturbative calculation might be justified.
}
since non-perturbative effects are 
suppressed due to the large top quark mass and
width~\cite{Fadin1,Fadin2} and since
the strong coupling is sufficiently small, $\alpha_s(C_F M_t \alpha_s)
\approx 0.15$.
The most common approach found in literature is based on numerical
solutions of a non-relativistic Schr\"odinger equation with a 
phenomenological QCD potential where the short-distance part originates
from loop calculations and the long-distance part is obtained
from fits to charmonium and bottomonium spectra~\cite{QCDpotential}. 
Although this approach
leads to applicable results, the use of a phenomenological potential
makes the systematic implementation of dynamic as well as 
kinematic relativistic corrections difficult, if not impossible.  
As a consequence all potential model predictions for the cross section
in the threshold region contain intrinsic uncertainties of relative
order $|\beta^2|\approx\alpha_s^2$. The actual size of these 
${\cal{O}}(\alpha_s^2)$ corrections\footnote{
Relativistic corrections to the non-relativistic cross section
actually result in a series in powers of $\alpha_s$ and $\beta$.
Because we consider the kinematic regime $|\beta|\lsim\alpha_s$, we
count powers of $\beta$ as powers of $\alpha_s$. For simplicity we
will call the ${\cal{O}}(\alpha_s^2)$, ${\cal{O}}(\alpha_s \beta)$ and
${\cal{O}}(\beta^2)$ relativistic corrections briefly 
``${\cal{O}}(\alpha_s^2)$ corrections''.
}
and their dependence on the 
c.m.~energy is unknown because a consistent and systematic determination
of these corrections has never been achieved.
It is the purpose of this letter to demonstrate how relativistic
perturbative ${\cal{O}}(\alpha_s^2)$ QCD corrections to the
non-relativistic stable heavy quark-antiquark cross section predicted
by the non-relativistic Schr\"odinger equation 
with a Coulombic QCD potential 
$V_{\mbox{\tiny QCD}}=-C_F\alpha_s/r$ and a fixed $\alpha_s$ can be
calculated. We want to emphasize that this work is not meant to
present all calculational details, but to show the main steps and
results of our calculations. A more detailed and explicit work will be
published later. We also would like to mention that at no point in
this work non-perturbative effects and electroweak corrections are
taken into account.
\par
The formula for the single photon mediated (i.e. vector-current
induced) non-relativistic cross section valid also for complex
energies reads~\cite{Hoang2}
\begin{equation}
R^{Q\bar Q, \mbox{\tiny NR}} \, = \,
\bigg[\,
\frac{\sigma(e^+e^-\stackrel{\gamma^*}{\longrightarrow}Q\bar Q)}
{\sigma_{pt}}
\,\bigg]_{\mbox{\tiny NR}}
\, = \,
\frac{3}{2}\,N_c\,e_Q^2\,\mbox{Im}\bigg\{\,
i\,\beta - C_F\,\alpha_s\,\bigg[\,
\gamma + \ln(-i\,\beta) + 
\Psi\Big(1-i\,\frac{C_F\,\alpha_s}{2\,\beta}\Big)
\,\bigg]
\,\bigg\}
\,,
\label{Rnonrel}
\end{equation}
where $\Psi$ is the digamma function,
$\Psi(z)\equiv\frac{d}{dz}\ln\Gamma(z)$,
$\sigma_{pt}=4\pi\alpha^2/(3s)$ and $N_c=3$ is the number of
colors. $e_Q$ is the electric heavy quark charge and $\alpha$ denotes
the fine structure constant. In the non-relativistic limit and for
unstable quarks, where the relation between the energy 
$E\equiv\sqrt{s}-2M_Q$ relative to the threshold point 
and the velocity can be approximated as
$\beta=[(E+i\,\Gamma_Q)/M_Q]^{1/2}$, $\Gamma_Q$ being the width of the
heavy quark, formula~(\ref{Rnonrel}) coincides with an expression
given in~\cite{Fadin1,Fadin2}.
For stable quarks and above threshold expression~(\ref{Rnonrel}) leads
to the famous Sommerfeld expression
\begin{equation}
R^{Q\bar Q, \mbox{\tiny NR}}_{\Gamma=0, \beta>0} \, = \,
\frac{3}{2}\,N_c\,e_Q^2\,
\frac{C_F\,\alpha_s\,\pi}
{1-\exp(-\frac{C_F\,\alpha_s\,\pi}{\beta})}
\,,
\label{Sommerfeldnonrel}
\end{equation}
whereas for stable quarks and below threshold eq.~(\ref{Rnonrel})
develops narrow resonances at the well known Coulomb energy
levels~\cite{Braun1, Hoang2}.
So far only ${\cal{O}}(\alpha_s)$ QCD corrections to the
non-relativistic cross section from the running of the strong
coupling~\cite{QCDpotential,Voloshin2,Yndurain1}  
and from short distances~\cite{Karplus1}  have been successfully
calculated.
The knowledge of the ${\cal{O}}(\alpha_s^2)$ corrections to 
expressions~(\ref{Rnonrel}) and (\ref{Sommerfeldnonrel}) is important
in order
to understand the structure and importance of relativistic corrections 
and might even lead to hints on the form of non-perturbative effects.
Because it is quite unclear how the ${\cal{O}}(\alpha_s^2)$ 
relativistic corrections to the non-relativistic cross section 
obtained from a Coulombic potential have to be implemented 
consistently into a potential model approach they will serve
as an order of magnitude estimate for the uncertainties inherent in 
potential model predictions.
\par
In this letter we explicitly determine the ${\cal{O}}(\alpha_s^2)$
corrections to the total single photon mediated heavy quark pair
production 
cross section from the relativistic energy-momentum relation, the 
relativistic phase space corrections and those ${\cal{O}}(\alpha_s^2)$
corrections which are multiplied by the SU(3) group theoretical factors
$C_F^2$ and $C_F T$ ($C_F=4/3$, $T=1/2$), where for the latter 
contribution only the effects from the heavy quark itself are taken into
account. We would like to stress that these corrections represent
a gauge invariant subset of all ${\cal{O}}(\alpha_s^2)$ QCD corrections
and that no model-like assumptions are imposed for our calculation.
The ${\cal{O}}(\alpha_s^2)$ QCD corrections involving the SU(3) group
theoretical factors $C_A C_F$ and $C_F T n_l$, where $C_A=3$ and
$n_l$ is the number of massless quark flavors, including the 
${\cal{O}}(\alpha_s^2)$ effects from the running of the strong 
coupling are not treated in this work, but can be determined along
the lines presented here. For the case of a non-negligible width 
of the heavy quark (as in $t\bar t$ production) further types of
corrections have to be taken into account. Apart from the perturbative
(multi-loop) corrections to the width of the free heavy quark also
corrections from the off-shellness of the decaying heavy quark, from
the interaction among the decay products and the other heavy quark (if
it is not decayed yet) and from time dilatation effects need to be
considered. Although the size and the interplay of all these effects
have been examined at various places in literature (see 
e.g.~\cite{Sumino1,Teubner1,Moedritsch1}) they are still not
completely understood yet as far as ${\cal{O}}(\alpha_s^2)$
corrections to the cross section are concerned. However, it is agreed
in~\cite{Sumino1,Teubner1,Moedritsch1} that the bulk of these effects
can be accounted for by using a momentum dependent width instead of a
constant one. This difficulty is completely ignored in this work. 
Therefore, all formulae presented here are strictly speaking only 
valid for stable quarks. However, during our calculations we never
assume that the squared velocity $\beta^2$ is a real number.
Therefore, our results could be easily implemented in a more complete
approach which treats all finite width effects properly. For now we
will use the naive replacement $E\to E+i\,\Gamma_Q$ in the spirit
of~\cite{Fadin1, Fadin2}, where $\Gamma_Q$ represents an appropriately
chosen constant heavy quark width, which is not necessarily equal to
the decay width of a free heavy quark. We think that this procedure is
justified in order to demonstrate the size of the
${\cal{O}}(\alpha_s^2)$  corrections calculated in this work in the
presence of a large quark width.
We finally would like to emphasize that 
relativistic corrections from the exchange of non-instantaneous
gluons (responsible for Lamb shift type corrections and real 
radiation effects for $s>4M_Q^2$) are of order
$\alpha_s^3$~\cite{Hoang1,Grinstein1} and beyond the level of accuracy
intended in this work.
\par
Let us start by reminding the reader that by means of the optical
theorem the non-relativistic cross section $R^{Q\bar Q, \mbox{\tiny
NR}}$, eq.~(\ref{Rnonrel}), can be written as~\cite{Fadin1,Fadin2}
\begin{equation}
R^{Q\bar Q, \mbox{\tiny NR}} \, = \,
\mbox{Im}\bigg[\,
N_c\,e_Q^2\,\frac{24\,\pi}{s}\,G_c(0,0)
\,\bigg]
\,,
\label{RGrelation}
\end{equation}
where $G_c$ is the Coulomb Green function satisfying the 
equation of motion 
\begin{equation}
\bigg[\,
-\frac{1}{M_Q}\vec{\nabla}^2_{\vec x} - \frac{a}{|{\vec x}|} - 
\frac{{\vec p}^{\,2}}{M_Q}
\,\bigg]\,G_c(\vec{x},\vec{y}) \, = \,
\delta^{(3)}(\vec{x}-\vec{y})
\,,\qquad
a \, \equiv \, C_F\,\alpha_s
\,.
\label{eqofmotion}
\end{equation} 
$\vec p$ is called the ``external three momentum'' for the rest of
this work. It should be noted that ${\vec p}^{\,2}/M_Q$ is equal to
the energy $E=\sqrt{s}-2M_Q$ only up to higher orders in $E$. In
quantum mechanics textbooks usually the convention is employed where
these higher order terms are set to zero. 
As explained later, we set ${\vec p}^{\,2}$ equal to the squared
(relativistic) three momentum of heavy quarks in the c.m.~frame.
An explicit analytic expression for $G_c$ in coordinate
space representation has to our
knowledge been calculated the first time in~\cite{Wichmann1}.
Our strategy is to calculate relativistic corrections to the 
Coulomb Green function $G_c$ by using (textbook quantum mechanics)
time independent perturbation theory (TIPT). This approach is
justified because only the exchange of instantaneous gluons needs to 
be taken into account. Using the 
concept of effective field theories the arising UV divergences are
removed by matching to recent two-loop results for the cross
section in the threshold region in the framework of QED~\cite{Hoang1}.
The cross section is then obtained via the optical theorem,
eq.~(\ref{RGrelation}).
Conceptually we follow the lines presented in~\cite{Hoang2}.
\par
The relativistic corrections to the Coulomb Green function arise from
three different sources: (a) the relativistic energy-momentum relation,
(b) $1/M_Q^2$ corrections to the Coulomb potential $V_{\mbox{\tiny QCD}}$
and (c) $1/M_Q^2$ corrections from the electromagnetic current which
produces and annihilates the quark-antiquark pair.
The corrections from the relativistic energy-momentum relation 
can be easily
determined by taking into account that $G_c$ can be written in the 
form~\cite{Fadin2}
\begin{equation}
G_c(\vec x,0) \, = \,
\int\!\!\frac{d^{\,3}{\vec p_0}}{(2\pi)^3}\,
e^{i{\vec p_0}{\vec x}}\,
\frac{M_Q}{{\vec p_0}^{\,2}-{\vec p}^{\,2}+i\epsilon}\,
\sum\limits_{m=0}^{\infty}\,
\prod\limits_{n=1}^{m}\,
\int\!\!\frac{d^{\,3}{\vec p_n}}{(2\pi)^3}\,
\frac{4\,\pi\,a}{({\vec p_{n-1}}-{\vec p_n})^2}\,
\frac{M_Q}{{\vec p_n}^{\,2}-{\vec p}^{\,2}+i\epsilon}
\,.
\label{GCoulomb}
\end{equation}
The relativistic expression for the particle-antiparticle propagation
$\int d{\vec p_n}^{\,3}/(2\pi)^3 M_Q/({\vec p_n}^{\,2}-
{\vec p}^{\,2}+i\epsilon)$
reads
\begin{equation}
-\,i\,\int\!\!\frac{d^{\,4} p_n}{(2\pi)^4}\,
S\Big(p_n+\Big(\sqrt{M_Q^2+{\vec p}^{\,2}},\vec 0\,\Big)\,\Big)\,
S\Big(p_n-\Big(\sqrt{M_Q^2+{\vec p}^{\,2}},\vec 0\,\Big)\,\Big)
\,,\qquad
p_n \, = \, (\,p_n^0,{\vec p_n}^{\,2}\,)
\,,
\end{equation}
where $S$ is the common Dirac propagator. Because the Coulomb
interaction is instantaneous the $p_n^0$ integration can be carried
out. An expansion in $1/M_Q^2$ then yields that the corrections
from the relativistic energy-momentum relation
can be implemented into expression~(\ref{GCoulomb}) by the replacement
\begin{equation}
\frac{M_Q}{{\vec p_n}^{\,2}-{\vec p}^{\,2}+i\epsilon}
\,\longrightarrow
\frac{M_Q}{{\vec p_n}^{\,2}-{\vec p}^{\,2}+i\epsilon}\,
\bigg[\,
1+\frac{{\vec p_n}^{\,2}+{\vec p}^{\,2}}{4\,M_Q^2}
\,\bigg]
\label{propagatorreplacement}
\end{equation}
for each particle-antiparticle propagator. As mentioned earlier,
the form of the correction factor on the r.h.s. of
eq.~(\ref{propagatorreplacement}) differs from the usual kinetic
energy correction used in quantum mechanics textbooks
because the relation between the external momentum $\vec{p}$ and the
c.m. energy reads
\begin{equation}
\frac{{\vec p}^{\,2}}{M_Q^2} \, \equiv \, \frac{\beta^2}{1-\beta^2}
\,.
\label{energymomentumfixing}
\end{equation}
For stable quarks definition~(\ref{energymomentumfixing}) leads to the
relation ${\vec p}^{\,2}/M_Q = E + E^2/(4M_Q) + {\cal{O}}(E^3)$
between the external momentum $\vec{p}$ and the energy $E$. Any other
definition of ${\vec p}^{\,2}$ would lead to a different form of the
r.h.s. of eq.~(\ref{propagatorreplacement}).
The final result for the cross section, of course, is independent of this
choice. We have chosen definition~(\ref{energymomentumfixing})
to facilitate our calculations.
The $1/M_Q^2$ corrections to the interaction potential are well known
and read
\begin{eqnarray}
\tilde V(\vec Q) & = &
-\frac{4\,\pi\,a}{{\vec Q^2}} +
\frac{\pi\,a}{M_Q^2} +
\frac{4\,\pi\,a}{M_Q^2}\,\bigg[\,
{\vec S_1}\,{\vec S_2} - 
\frac{({\vec Q}\,{\vec S_1})\,({\vec Q}\,{\vec S_2})}{{\vec Q}^2}
\,\bigg]
\nonumber\\[2mm] & &  -
\frac{4\,\pi\,a}{M_Q^2}\,\bigg[\,
\frac{{\vec p}^{\,2}}{{\vec Q}^2}  - 
\frac{({\vec Q}\,{\vec p}\,)^2}{{\vec Q}^4}
\,\bigg] -
i\,\frac{6\,\pi\,a}{M_Q^2} \,\bigg[\, 
 ({\vec S_1}+{\vec S_2})\,\frac{{\vec Q}\times{\vec p}}{{\vec Q}^2}
\,\bigg]
\label{Potential1}
\end{eqnarray}
in momentum space representation, where $\vec Q$ is the (three) 
momentum flowing through the gluons and the $\vec S_{1/2}$ represent
the quark/antiquark spin operators. In eq.~(\ref{Potential1}) the 
Coulomb interaction is also displayed. 
The $1/M_Q^2$ corrections from the electromagnetic current which
produces and annihilates the heavy quark-antiquark pair lead to the 
insertion of the factor  $\{1-\frac{{\vec p_0}^{\,2}}{3
M_Q^2}[\frac{3}{4}+ {\vec S_1}{\vec S_2}]\}$ 
into expression~(\ref{GCoulomb}).
\par
Taking into account that the $Q\bar Q$ pair is produced in a
($J^{PC}=1^{-\,-}$) ${}^3S_1$ state\footnote{
The production of a ${}^3D_1$ state is proportional to the modulus
squared of the second derivative of the heavy quark-antiquark wave
function at the origin and therefore suppressed by $|\beta|^4\approx
\alpha_s^4$. This is beyond the intended accuracy. 
}
the relativistic corrections to the Coulomb Green 
function can be rewritten in terms of corrections induced by an 
effective interaction potential which in coordinate space representation
takes the simple form 
\begin{eqnarray}
V_{3S1}({\vec x}) & = & -\frac{a}{|{\vec x}|}\,
 \bigg[\,1+\frac{3\,{\vec p}^{\,2}}{2\,M_Q^2}\,\bigg] +
\frac{11}{3}\,\frac{\pi\,a}{M_Q^2}\,\delta^{(3)}(\vec x) -
\frac{5}{4\,M_Q}\,\frac{a^2}{|{\vec x}|^2}
\label{Potential2}
\,,
\end{eqnarray}
where the Coulomb potential is also displayed.
In addition, there remains a correction to the Coulomb Green function,
which cannot be expressed in terms of an interaction potential. This
correction takes the form
\begin{equation}
\delta G_c(0,0) \, = \,
-\lim\limits_{|{\vec x}|\to 0} \bigg[\,\frac{1}{M_Q^2}\,\bigg(\,
\frac{7}{6}\,{\vec\nabla}^2+ {\vec p}^{\,2}
\,\bigg)\,G_c(\vec x,0)
\,\bigg]
\label{deltaphase}
\end{equation}
and essentially represents relativistic phase space corrections.
\par
It is now straightforward to determine all 
${\cal{O}}(\alpha_s^2)$ corrections to the
Coulomb Green function. The corrections from the first term on the
r.h.s. of (\ref{Potential2}), called kinetic energy corrections
later in this paper,  can be trivially implemented by the 
replacement $a\to a(1+3{\vec p}^{\,2}/2 M_Q^2)$ in $G_c$. The corrections
from the second and third term, later called dynamical corrections, 
can be calculated via coordinate space TIPT. The phase space
corrections of eq.~(\ref{deltaphase}) can be evaluated by employing the 
equation of motion (\ref{eqofmotion}). The arising UV (short-distance) 
divergences can
be regularized by considering corrections to $G_c(\vec x,0)$ and taking
the limit $|\vec x|\to 0$ afterwards\footnote{
We emphasize that this method is not claimed to be a consistent way of
UV regularization in coordinate space. However, in our case a quite
sloppy treatment of UV divergences is allowed because we later match
our result directly to the two-loop expression for the cross section
in the threshold region. (See also~\cite{Hoang2}.)
}.
Taking into account that the relation between the cross section
$R^{Q\bar Q}$ and the Green function, eq.~(\ref{RGrelation}), leads to 
another $1/M_Q^2$ phase space correction and using 
relation~(\ref{energymomentumfixing}) the result for the cross section
reads
\begin{equation}
R^{Q\bar Q} \, = \,\frac{3}{2}\,N_c\,e_Q^2\,a\,
\mbox{Im}\Big[\,H_a(a,\beta)\,\Big]\,
\bigg\{\,
1 + a\,\Big[\,div\,\Big] + a^2\,\Big[\,div\,\Big] + 
\frac{2}{3}\,a^2\,\mbox{Re}\Big[\,H_a(a,\beta)\,\Big]
\,\bigg\}
\label{Runrenormalized}
\,,
\end{equation}
where
\begin{equation}
\label{Hdef}
H_a(a,\beta) \, \equiv \, \Big(1-\frac{1}{3}\,\beta^2\Big)\,
\bigg\{\,
 i\,\frac{\beta}{a} - \,(1+\beta^2)\,
\bigg[\,
\gamma + \ln(-i\,\beta) + 
\Psi\Big(1-i\, a\,\frac{1+\beta^2}{2\,\beta}\Big)
\,\bigg]
\,\bigg\}
\,.
\end{equation}
It should be noted that the term $\ln(-i\,\beta)$ in the function
$H_a$ does not lead to a singular behavior of $H_a$ in the limit
$\beta\to 0$ because this logarithm is cancelled by a corresponding
logarithmic term generated by the digamma function in the same
limit. It is a remarkable fact that in the framework of an expansion
in Feynman diagrams the combination 
$\gamma + \Psi(1-i\, a \frac{1+\beta^2}{2\,\beta})$ is generated
entirely by diagrams of higher order than the diagrams which produce
the explicit term $\ln(-i\,\beta)$. (See also~\cite{Hoang2}.)
The terms in eq.~(\ref{Runrenormalized}) symbolized by $[div]$
represent divergent and $\beta$-independent contributions.   
The divergences originate from the
integration region $|{\vec x}|\to 0$ in TIPT and have to be considered
as UV divergences which indicate that the electromagnetic 
current which produces and annihilates the heavy quark pair has to be
renormalized. This renormalization is usually achieved by the
determination of the corresponding counterterm via matching
to amplitudes calculated in covariant (multi-loop) perturbation
theory in the framework of QCD~\cite{Lepage1,Bodwin1}. 
Fortunately this lengthy procedure
is not necessary in our case because expression~(\ref{Runrenormalized})
can be matched directly to a recent two-loop calculation of the
fermion-antifermion cross section in the threshold region in the
framework of QED~\cite{Hoang1} which is sufficient to determine the
${\cal{O}}(\alpha_s^2)$ corrections we are interested in. This 
``direct matching'' procedure~\cite{Hoang2} is carried out in the formal
limit $a\ll\beta\ll 1$ for stable quarks, where predictions in the
non-relativistic effective theory and conventional multi-loop
perturbation theory in QCD have to coincide. Expanding up to 
next-to-next-to-leading order in $\beta$ 
(and including only the ${\cal{O}}(\alpha_s^2)$ contributions
with the color factors $C_F^2$ and $C_F T$ we are interested in) 
the two-loop expression (i.e. including Born,
one-loop and two-loop contributions) for the cross section
reads~\cite{Hoang1}
\begin{eqnarray}
R^{Q\bar Q}_{\mbox{\tiny 2 Loop}} & = &
N_c\,e_Q^2\,\bigg\{\,\bigg[\,
\frac{3}{2}\,\beta-\frac{1}{2}\,\beta^3
\,\bigg] +
\frac{C_F\,\alpha_s}{\pi}\,\bigg[\,
\frac{3\,\pi^2}{4}-6\,\beta+\frac{\pi^2}{2}\,\beta^2
\,\bigg]
\nonumber\\[2mm] & & \qquad
+\, \alpha_s^2\,\bigg[\,
\frac{C_F^2\,\pi^2}{8\,\beta} - 3\,C_F^2 + 
\bigg(\, \frac{5\,C_F^2\,\pi^2}{24} + \frac{3}{2}\,C_2 - C_F^2\,\ln\beta 
\,\bigg)\,\beta
\,\bigg]
\,\bigg\}
\,,
\label{Rtwoloop}
\end{eqnarray}
where
\begin{eqnarray}
C_2 & \equiv &
C_F^2\,\bigg[\, \frac{1}{\pi^2}\,\bigg(\,
\frac{39}{4}-\zeta_3
\,\bigg) +
\frac{4}{3}\,\ln 2 - \frac{35}{18}
\,\bigg] + 
C_F\,T\,\bigg[\,
\frac{4}{9}\,\bigg(\, \frac{11}{\pi^2} - 1
\,\bigg)
\,\bigg]
\,.
\label{C2def}
\end{eqnarray}
Expanding eq.~(\ref{Runrenormalized}) in the same way and demanding
equality to expression~(\ref{Rtwoloop}) the divergent contributions 
in eq.~(\ref{Runrenormalized}) are  unambiguously removed and replaced
by constant (i.e. $\beta$-independent)
terms. The final result for the cross section then reads
\begin{equation}
R^{Q\bar Q} \, = \, \frac{3}{2}\,N_c\,e_Q^2\,C_F\,\alpha_s\,
\mbox{Im}\Big[\,H_a(C_F\,\alpha_s,\beta)\,\Big]\,
\bigg\{\,
1 - 4\,C_F\,\frac{\alpha_s}{\pi} + \alpha_s^2\,C_2 + 
\frac{2}{3}\,C_F^2\,\alpha_s^2\,
       \mbox{Re}\Big[\,H_a(C_F\,\alpha_s,\beta)\,\Big]
\,\bigg\}
\,.\,
\label{Rrenormalized}
\end{equation}
In eq.~(\ref{Rrenormalized}) the well known ${\cal{O}}(C_F\alpha_s)$
short-distance correction $-4 C_F\alpha_s/\pi$~\cite{Karplus1} is
successfully recovered. It is an
interesting fact that the size of the contribution from $C_2=-0.24$ is
an order of magnitude smaller than the one of the function $H_a$. 
For convenience, $C_2$ will be called ${\cal{O}}(\alpha_s^2)$
short-distance correction in the following discussion.
However, we would like to stress that a unique identification
of short-distance and long-distance contributions in 
the combination $\alpha_s^2 C_2 + \frac{2}{3}\, C_F^2 \alpha_s^2
\mbox{Re}[H_a]$ is impossible because such a procedure is
cut-off-dependent. In the language of conventional perturbation theory
(in the number of loops), expression~(\ref{Rrenormalized}) resums all
contributions of order 
$\beta(C_F\alpha_s/\beta)^n\times[1, C_F\alpha_s, \beta^2,
C_F\alpha_s\beta, C_F^2\alpha_s^2, C_F T\alpha_s^2]$, 
$n=0,1,2,\ldots, \infty$, in an expansion for small $\beta$.
Because expression~(\ref{Rrenormalized}) is also valid for complex
energies it is applicable for $t\bar t$ production, where
the large top width has to be taken into account\footnote{
For complex energies there is an ambiguity in the definition
of the function $H_a$ of order $\beta^2$ which cannot be removed
by the matching to the two-loop result calculated for stable
quarks. For the case of unstable quarks this ambiguity amounts to
$\mbox{Im}[a\beta^2]\sim C_F\alpha_s \Gamma_Q/M_Q$ in the total cross
section, where $\Gamma_Q$ is the quark width. For the case of $t\bar t$
production ($\Gamma_t \approx 1.5$~GeV) this ambiguity is of order
$0.1\%$ which is beyond the intended accuracy. For bottom and charm
quarks this ambiguity can be ignored.
}.
\par
\begin{figure}[t]
\begin{center}
\leavevmode
\epsfxsize=4cm
\epsffile[220 420 420 550]{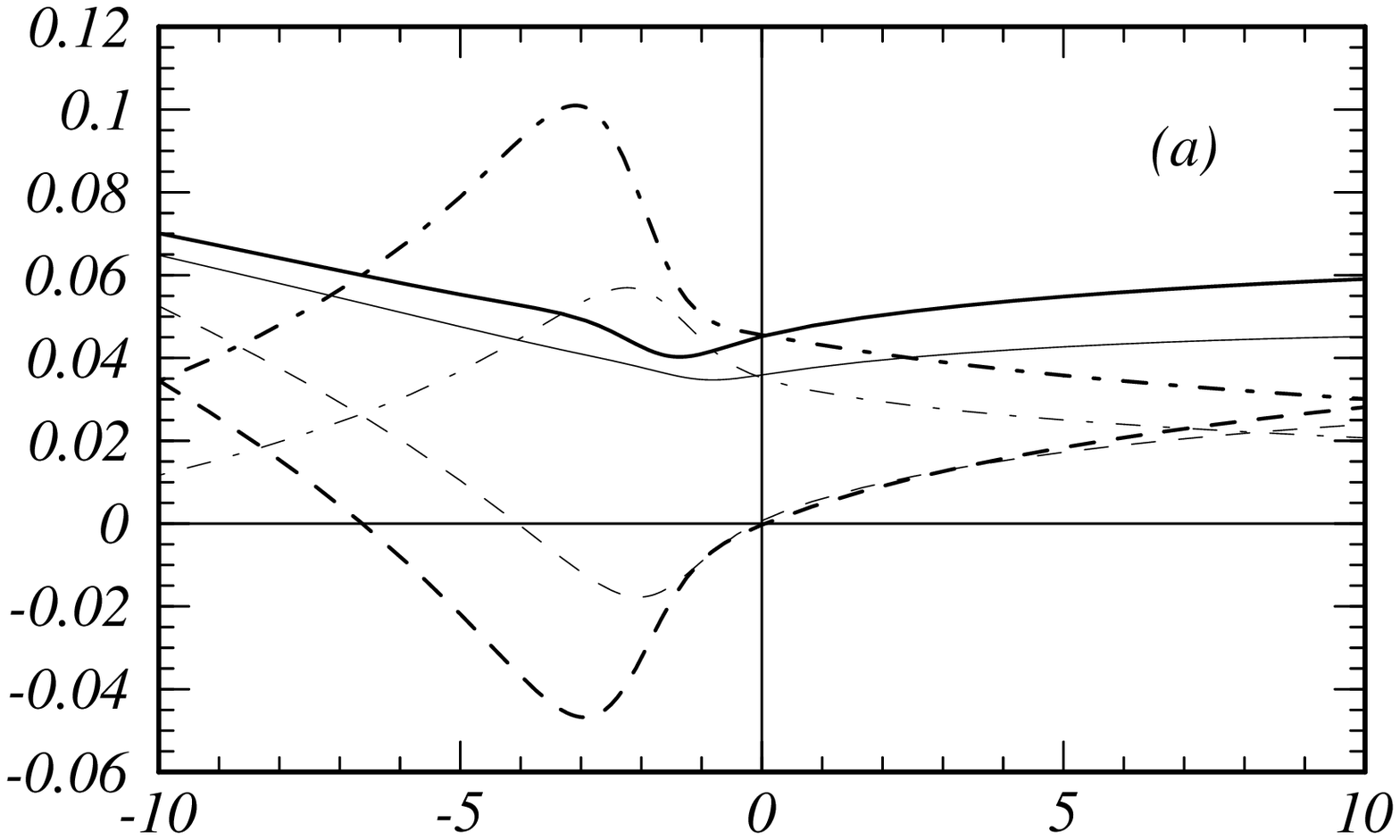}\\
\vskip 3.3cm
\epsfxsize=4cm
\leavevmode
\epsffile[220 420 420 550]{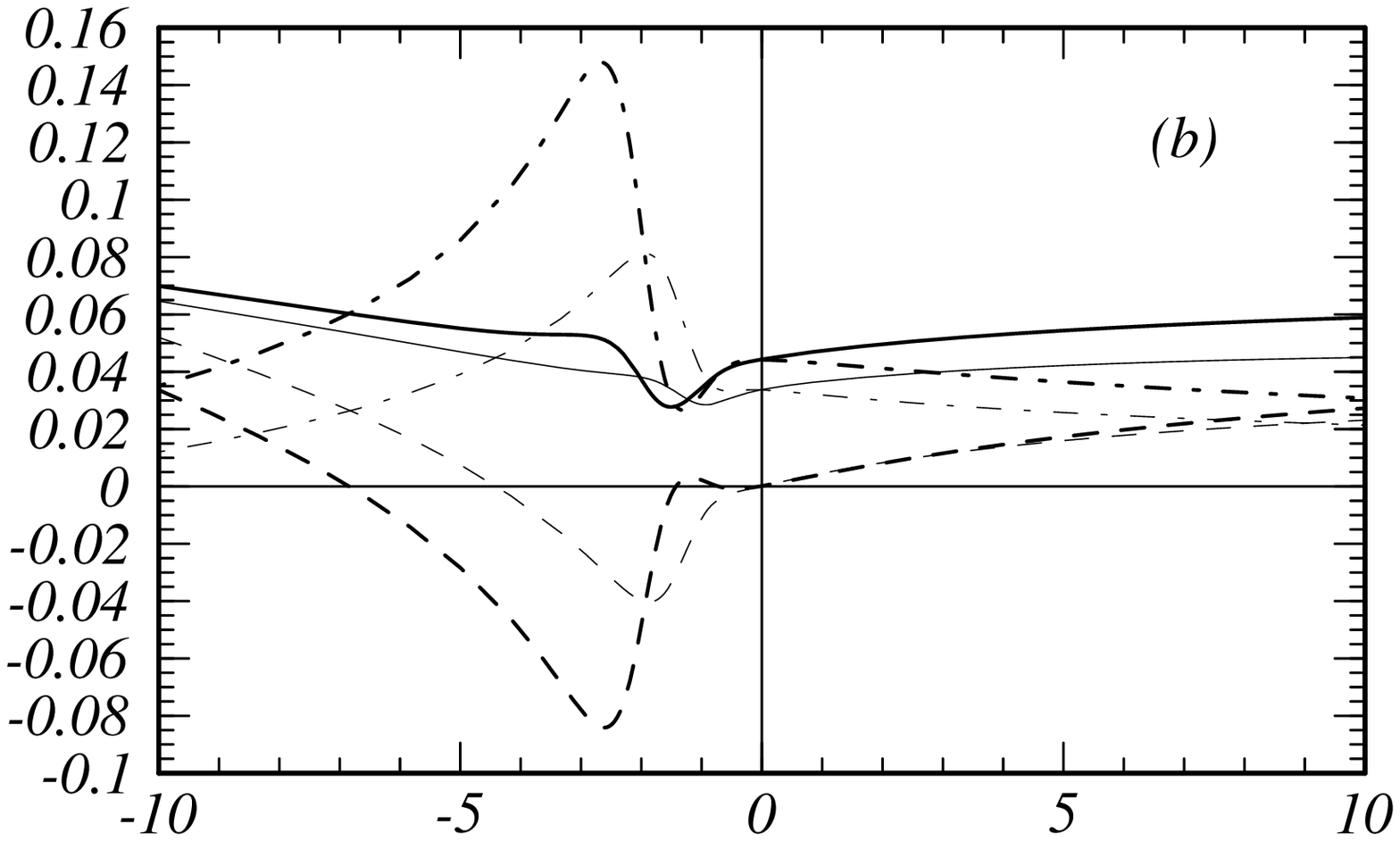}
\vskip  -46mm  $\displaystyle{\mbox{\hspace{4.3cm}}
                                \bf M_t=175\,\mbox{GeV}}$\\
\vskip  1mm  $\displaystyle{\mbox{\hspace{4.3cm}}
                                \bf \Gamma_t=1.55\,\mbox{GeV}}$
\vskip  0mm  $\displaystyle{\mbox{\hspace{12cm}}\bf E
(\mbox{GeV})}$
\vskip  40mm  $\displaystyle{\mbox{\hspace{4.3cm}}
                                \bf M_t=175\,\mbox{GeV}}$\\
\vskip  1mm  $\displaystyle{\mbox{\hspace{4.3cm}}
                                \bf \Gamma_t=0.80\,\mbox{GeV}}$
\vskip  2mm  $\displaystyle{\mbox{\hspace{12cm}}\bf E
(\mbox{GeV})}$
\vskip  1mm
 \caption{\label{fig1} 
 The relative ${\cal{O}}(\alpha_s^2)$ corrections to the
 total non-relativistic $t\bar t$ production cross section
 $\Delta^1$ (dashed lines), $\Delta^2$ (dashed-dotted lines) and
 $\Delta^1+\Delta^2$ (solid lines) for $\alpha_s=0.13$ (thin lines)
 and $\alpha_s=0.16$ (thick lined) as described in the text.
 }
 \end{center}
\end{figure}
Although the ${\cal{O}}(C_A C_F\alpha_s^2)$ and 
${\cal{O}}(C_F T n_l\alpha_s^2)$ corrections to the heavy 
quark-antiquark cross section in the threshold region are still unknown,
it is instructive to examine the ${\cal{O}}(\alpha_s^2)$ corrections
contained in eq.~(\ref{Rrenormalized}) 
for the case of $t\bar t$ production.
In Fig.~\ref{fig1} the sum of the relative ${\cal{O}}(\alpha_s^2)$
kinetic energy and the phase space corrections,  
$\Delta^1\equiv(\frac{2}{3} N_c e_Q^2 C_F \alpha_s
\mbox{Im}[H_a]-R^{Q\bar Q,  
\mbox{\tiny NR}})/R^{Q\bar Q, \mbox{\tiny NR}}$ (dashed lines), 
the ${\cal{O}}(\alpha_s^2)$ dynamical 
corrections including the ${\cal{O}}(\alpha_s^2)$ short-distance
contribution,
$\Delta^2\equiv\alpha_s^2 C_2 + \frac{2}{3}\, C_F^2 \alpha_s^2
\mbox{Re}[H_a]$ (dashed-dotted lines),
and their sum, $\Delta^1+\Delta^2$ (solid lines), are plotted for
$\alpha_s=0.13$ (thin lines) and $\alpha_s=0.16$ (thick lined) in the
energy range $-10$~GeV$\,< E < 10$~GeV, where $E = \sqrt{s}-2M_t$
and $M_t=175$~GeV.
As mentioned earlier the top decay width is implemented by the naive
replacement $E\to E+i\,\Gamma_t$, which leads to the relation
$\beta=[1-4 M_t^2/(E+i \Gamma_t+2 M_t)^2]^{1/2}$ between the velocity
$\beta$ and the energy $E$. In Fig.~\ref{fig1}(a) $\Gamma_t=1.55$~GeV,
whereas in  Fig.~\ref{fig1}(b) we have chosen $\Gamma_t=0.80$~GeV.
It is striking that the strong energy dependence of $\Delta^1$ and 
$\Delta^2$ around the $1S$ peak is cancelled in their sum leaving a 
fairly stable correction between $3\%$ and $7\%$ over the whole
threshold region\footnote{
We would like to mention that $\Delta^2$ contains contributions of
order $\alpha_s^2 \beta^2$ coming from the $\beta^2$ terms in the
function $H_a$. These terms represent contributions beyond the
intended accuracy and are not included in our analysis. The size of
these contributions to $\Delta^2$ does not exceed $0.5\%$ in the
considered energy range.
}. 
This shows that the $1S$~peak, which is the most important
characteristic of the total $t\bar t$ production cross section in the
threshold region~\cite{QCDpotential}, is barely shifted by the
${\cal{O}}(\alpha_s^2)$ 
corrections determined in this work. (See also the near cancellation
of the ${\cal{O}}(\alpha_s^4)$ contributions to the energy
levels $E_n$ in eq.~(\ref{energylevels}) for $n=1$.) In particular,
the sum $\Delta^1+\Delta^2$ is 
fairly insensitive to variations in the choice
of the value of $\Gamma_t$ indicating that the size of the
${\cal{O}}(\alpha_s^2)$ corrections calculated in this work is not
affected by our ignorance of a consistent treatment of all finite
width effects. The variation of the size of $\Delta^1+\Delta^2$ for
different choices of $\alpha_s$ further shows that once all
${\cal{O}}(\alpha_s^2)$ corrections are calculated the remaining
relative theoretical uncertainty for the total cross section can be
expected
at the level of $1\%$. Taking the size of $\Delta^1+\Delta^2$ as an
order of magnitude estimate for the sum of all ${\cal{O}}(\alpha_s^2)$
corrections and because a consistent ${\cal{O}}(\alpha_s^2)$ 
analysis has never been accomplished in the framework of potential
models for $t\bar t$ production we come to the conclusion that an
uncertainty of the order $5\%$ percent is contained in all
predictions for the total $t\bar t$ production cross section
based on phenomenological potentials. 
\par
\begin{figure}[t]
\begin{center}
\leavevmode
\epsfxsize=5cm
\epsffile[220 420 420 550]{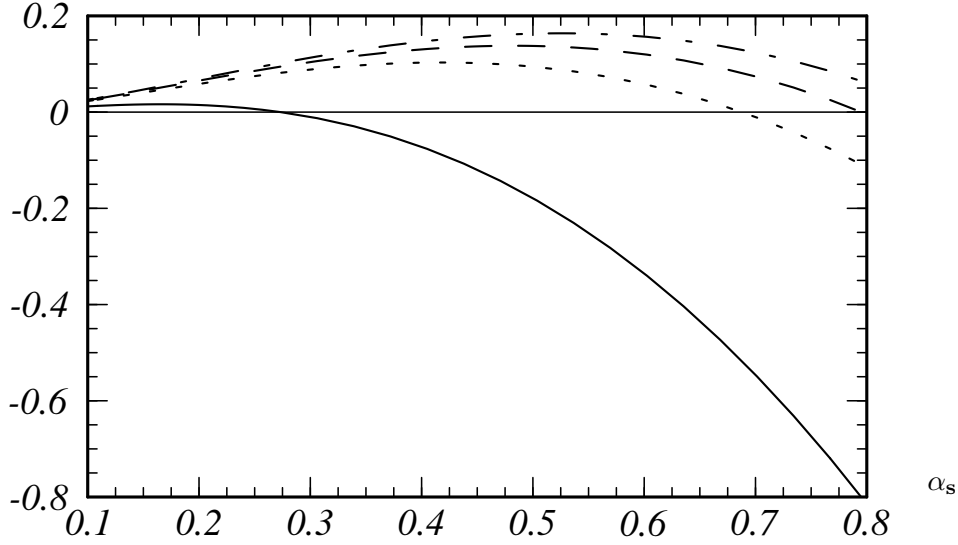}\mbox{\hspace{4mm}}\\
\vskip 30mm  $\displaystyle{\mbox{\hspace{12cm}}\bf \alpha_s}$
\vskip 6mm
 \caption{\label{fig2} 
 The relative ${\cal{O}}(\alpha_s^2)$ corrections 
 $\delta_n$ for $n=1$~(solid line), $2$~(dashed line),  
 $3$~(dashed-dotted line) and $\infty$~(dotted line) for values 
 of $\alpha_s$ in the range $0.1 < \alpha_s < 0.8$
 as described in the text.
 }
 \end{center}
\end{figure}
Evaluating formula~(\ref{Rrenormalized}) for stable quarks above
threshold we obtain
\begin{eqnarray} 
R^{Q\bar Q}_{\Gamma=0,\beta>0} &  = & 
\frac{3}{2}\,N_c\,e_Q^2\,
\beta\,\Big(\,1-\frac{1}{3}\beta^2\,\Big)\,
\frac{(\frac{2C_F\alpha_s\pi}{v_{rel}})}
{1-\exp(-\frac{2C_F\alpha_s\pi}{v_{rel}})}\,
\bigg\{\,
1-4\,C_F\,\frac{\alpha_s}{\pi} 
\nonumber \\[2mm] & & \mbox{\hspace{2cm}}
 + \alpha_s^2\,\bigg[\,C_2 -
\frac{2}{3}\,C_F^2\,\bigg(\,
\gamma + \ln\beta + 
\mbox{Re}\Psi\Big(1-i\,\frac{C_F\,\alpha_s}{2\,\beta}\Big)
\,\bigg)
\,\bigg]
\,\bigg\}
\,,
\label{Sommerfeldrelat}
\end{eqnarray}
where $v_{rel} = 2\beta/(1+\beta^2)$ is the relativistic 
relative velocity of the produced quark pair. This verifies a
suggestion for the form of the relativistic extension of the
Sommerfeld expression made in~\cite{Hoang1}. For $\beta\to 0$ this leads
to the finite expression
\begin{equation}
R^{Q\bar Q}_{\Gamma=0,\beta=0} \, = \,
\frac{3}{2}N_c\,e_Q^2\,C_F\,\alpha_s\,\pi\,
   \bigg\{\,1-4\,C_F\,\frac{\alpha_s}{\pi} +
\alpha_s^2\,\bigg[\, C_2 - 
\frac{2}{3}\,C_F^2\,\bigg(\,
\ln\Big(\frac{C_F\,\alpha_s}{2}\Big)+\gamma
\,\bigg)
\,\bigg]
\,\bigg\}
\,.
\end{equation}
For stable quarks and below threshold 
formula~(\ref{Rrenormalized}) develops narrow resonances at
the spin triplet ($n{}^3S_1$) energy levels
($n=1,2,\dots,\infty$)\footnote{
The derivation of the energy levels in eq.~(\ref{energylevels}) from
expression~(\ref{Rrenormalized}) is carried out according to
eqs.~(38)--(40) in~\cite{Hoang2}. We also would like to note that
the ${\cal{O}}(\alpha_s^4)$ contributions to the
$n{}^3S_1$ energy levels in eq.~(\ref{energylevels}) are consistent
with the effective potential $V_{3S1}$ given in
eq.~(\ref{Potential2}). 
}
\begin{equation}
E_n \, = \, 
-\frac{M_Q\,C_F^2\,\alpha_s^2}{4\,n^2} +
\frac{M_Q\,C_F^4\,\alpha_s^4}{4\,n^4}\,\bigg[\,
\frac{11}{16} - \frac{2}{3}\,n
\,\bigg]
\,.
\label{energylevels}
\end{equation}
Parameterizing the resonances in the form ($\alpha=1/137$) 
\begin{equation}
R^{Q\bar Q}_{\Gamma=0,-i\beta>0} \, = \,
\frac{9\,\pi}{\alpha^2}\,\sum\limits_{n=1}^\infty\,
\Gamma(n\to e^+e^-)\,M_n\,\delta(s-M_n^2)
\,,
\end{equation}
where the $M_n\equiv 2M_Q+E_n$ are the vector resonance masses,
we can extract the corrections to the leptonic widths 
($n=1,2,\ldots,\infty$), 
\begin{eqnarray}
\Gamma(n\to e^+e^-) & = &
N_c\,e_Q^2\,
\frac{16\,\pi\,\alpha^2}{3}\,\frac{|\Psi_n^c(0)|^2}{M_n^2}\,
\bigg\{\,
1-4\,C_F\,\frac{\alpha_s}{\pi} + \delta_n
\,\bigg\}
\,,\\[2mm] 
\delta_n & = & 
\alpha_s^2\,\bigg[\,
C_2 - \frac{2}{3}\,C_F^2\,\bigg(\,
\frac{5}{2\,n^2} + \ln\Big(\frac{C_F\,\alpha_s}{2\,n}\Big) -
\frac{1}{n} + \Psi(n) + \gamma
\,\bigg)
\,\bigg]
\,.
\end{eqnarray}
$|\Psi_n^c(0)|^2=M_Q^3 C_F^3 \alpha_s^3 /8\pi n^3$ is the modulus
squared of the (unperturbed) Coulomb wave function at the origin with
the radial quantum number $n$. It should be noted that the relative 
${\cal{O}}(\alpha_s^2)$ correction $\delta_n$ remains finite in the
limit $n\to\infty$ (called ``small binding limit'' in~\cite{Barbieri1}),
\begin{equation}
\delta_\infty \, \equiv \,
\lim\limits_{n\to\infty}\,\delta_n \, = \,
\alpha_s^2\,\bigg[\,
C_2 - \frac{2}{3}\,C_F^2\,\bigg(\,
\ln\Big(\frac{C_F\,\alpha_s}{2}\Big) + \gamma
\,\bigg)
\,\bigg]
\,.
\end{equation}
In Fig.~\ref{fig2} 
the relative ${\cal{O}}(\alpha_s^2)$ correction
$\delta_n$ is plotted for $n=1$~(solid line), $2$~(dashed line),  
$3$~(dashed-dotted line) and $\infty$~(dotted line) for values 
of $\alpha_s$ in the range $0.1 < \alpha_s < 0.8$.
\par
I am grateful to B. Grinstein, J.H. K\"uhn, A. Manohar and T. Teubner
for useful discussions and comments.
\newpage
\sloppy
\raggedright
\def\app#1#2#3{{\it Act. Phys. Pol. }{\bf B #1} (#2) #3}
\def\apa#1#2#3{{\it Act. Phys. Austr.}{\bf #1} (#2) #3}
\def\lhc{Proc. LHC Workshop, CERN 90-10}
\def\npb#1#2#3{{\it Nucl. Phys. }{\bf B #1} (#2) #3}
\def\nP#1#2#3{{\it Nucl. Phys. }{\bf #1} (#2) #3}
\def\plb#1#2#3{{\it Phys. Lett. }{\bf B #1} (#2) #3}
\def\prd#1#2#3{{\it Phys. Rev. }{\bf D #1} (#2) #3}
\def\pra#1#2#3{{\it Phys. Rev. }{\bf A #1} (#2) #3}
\def\pR#1#2#3{{\it Phys. Rev. }{\bf #1} (#2) #3}
\def\prl#1#2#3{{\it Phys. Rev. Lett. }{\bf #1} (#2) #3}
\def\prc#1#2#3{{\it Phys. Reports }{\bf #1} (#2) #3}
\def\cpc#1#2#3{{\it Comp. Phys. Commun. }{\bf #1} (#2) #3}
\def\nim#1#2#3{{\it Nucl. Inst. Meth. }{\bf #1} (#2) #3}
\def\pr#1#2#3{{\it Phys. Reports }{\bf #1} (#2) #3}
\def\sovnp#1#2#3{{\it Sov. J. Nucl. Phys. }{\bf #1} (#2) #3}
\def\sovpJ#1#2#3{{\it Sov. Phys. LETP Lett. }{\bf #1} (#2) #3}
\def\jl#1#2#3{{\it JETP Lett. }{\bf #1} (#2) #3}
\def\jet#1#2#3{{\it JETP Lett. }{\bf #1} (#2) #3}
\def\zpc#1#2#3{{\it Z. Phys. }{\bf C #1} (#2) #3}
\def\ptp#1#2#3{{\it Prog.~Theor.~Phys.~}{\bf #1} (#2) #3}
\def\nca#1#2#3{{\it Nuovo~Cim.~}{\bf #1A} (#2) #3}
\def\ap#1#2#3{{\it Ann. Phys. }{\bf #1} (#2) #3}
\def\hpa#1#2#3{{\it Helv. Phys. Acta }{\bf #1} (#2) #3}
\def\ijmpA#1#2#3{{\it Int. J. Mod. Phys. }{\bf A #1} (#2) #3}
\def\ZETF#1#2#3{{\it Zh. Eksp. Teor. Fiz. }{\bf #1} (#2) #3}
\def\jmp#1#2#3{{\it J. Math. Phys. }{\bf #1} (#2) #3}
\def\yf#1#2#3{{\it Yad. Fiz. }{\bf #1} (#2) #3}

\end{document}